\documentclass[authoryear,preprint]{elsarticle}

\usepackage{graphicx}
\usepackage{epsfig}
\usepackage{amsmath}
\usepackage{amssymb}
\usepackage{amsfonts}

\journal{Advances in Space Research}

\begin{document}

\begin{frontmatter}

\title{Pioneer 10 Doppler data analysis: disentangling periodic and secular anomalies }

\author{A. Levy and B. Christophe}
\address{ONERA/DMPH, 29 av. Division Leclerc, F-92322 Chatillon, France}
\ead{bruno.christophe@onera.fr}

\author{P. B\'erio and G. M\'etris}
\address{Geoscience Azur, Universit\'e Nice Sophia-Antipolis, OCA, Avenue Copernic F-06130 Grasse, France}
\ead{gilles.metris@obs-azur.fr}

\author{J-M. Courty and S. Reynaud}
\address{LKB, UPMC, case 74, CNRS, ENS, F-75252 Paris cedex 05, France}
\ead{courty@spectro.jussieu.fr}

\begin{abstract}
This paper reports the results of an analysis of the Doppler tracking data
of Pioneer probes which did show an anomalous behaviour. 
A software has been developed for the sake of performing a data analysis as independent as possible
from that of J.~Anderson et al. \citep{anderson}, using the same data set. 
A first output of this new analysis is a confirmation of the existence of a secular anomaly 
with an amplitude about 0.8\,nms$^{-2}$ compatible with that reported by Anderson et al. 
A second output is the study of periodic variations of the anomaly, which we characterize as functions of 
the azimuthal angle $\varphi$ defined by the directions Sun-Earth Antenna and Sun-Pioneer. 
An improved fit is obtained with periodic variations written as the sum of a secular acceleration and 
two sinusoids of the angles $\varphi$ and $2\varphi$.
The tests which have been performed for assessing the robustness of these results are presented. 
\end{abstract}


\end{frontmatter}

\parindent=0.5 cm

\section{Introduction} 
The Pioneer 10 and 11 spacecrafts performed accurate celestial mechanics experiment after their encounters with Jupiter and Saturn. 
Doppler tracking by the NASA Deep Space Network \citep{moyer} of the two spacecrafts 
which followed escape orbits to opposite ends of the solar system revealed an anomalous behaviour which remains imperfectly understood. 
Precisely, the tracking data do not meet the expectations drawn from standard gravity 
force law and a better fit is obtained by adding a constant Sun-ward acceleration \citep{andersonPRL}. 
This anomalous acceleration, now refered to as Pioneer anomaly, has a similar magnitude for the two spacecrafts. The value reported by the JPL team who discovered the anomaly is 0.874$\pm$0.133\,nms$^{-2}$ \citep{anderson}. 

The extensive analysis by the JPL team has been published after years of cross checks~\citep{anderson}.
The presence of an anomaly and its magnitude have been confirmed by different analysis software \citep{markwardt,olsen}. 
A number of mechanisms have been considered as attempts of explanations of the anomaly as a systematic 
effect generated  by the spacecraft itself or its environment (see as an example \citep{nieto}) 
but they have not led to a satisfactory understanding to date.
If confirmed, the Pioneer signal might reveal an anomalous behaviour of gravity 
at scales of the order of the size of the solar system and thus have a 
strong impact on fundamental physics, astrophysics and cosmology.
It is therefore important to explore all its facets in the wider context
of navigation and test of the gravity law in the solar system 
(see for example~\citep{reynaud,bertolami} and references therein).

Crucial informations are expected to come out from the re-analysis of Pioneer data 
\citep{nietodata,turyshevdata}.
An international collaboration has been built to this aim
within the frame of International Space Science Institute \citep{issi}. 
The Pioneer data which had been analysed by the Anderson team \citep{anderson}
have been made available by Slava Turyshev (JPL) in the framework of this collaboration.
They consist in Orbit Data Files (ODF) which contain in particular Pioneer 10 Doppler data 
from November 30th, 1986 to July 20th, 1998.

The aim of the present paper is to report some results of the analysis of these data
performed by a collaboration between three groups at ONERA, OCA and LKB within the ``Groupe Anomalie Pioneer'' \citep{gap}. 
A dedicated software called ODYSSEY has been developed to this purpose. 
The first result is to confirm the existence and magnitude of the anomalous
secular acceleration reported by Anderson et al., using different and as independent as possible tools. 
The main motivation of the present paper is to study the periodic variations of the anomaly,
which are known to exist besides the secular Pioneer anomaly. 

The presence of diurnal and seasonal variations in the residuals has been discussed by Anderson et al. \citep{anderson}. 
We may emphasize that such modulated anomalies are unlikely to be due to anything related to the spacecraft
or its environment. 
Anderson et al. \citep{anderson} have proposed that they are due to modeling errors such as errors in the 
Earth's ephemeris, the orientation of the Earth's spin axis or the station's coordinates.  However, these parameters are strongly constrained by other observational methods and it seems difficult to change them enough to explain the periodic anomaly. We will show here that this periodic anomaly can be at least partly represented in terms of a modulation  of the Doppler signal as a function of a unique azimuthal angle having a physical meaning.

\section{Development of the ODYSSEY software}

The Pioneer 10/11 spacecrafts were tracked by the Deep Space Network (DSN) antennas. 
A S-Band signal at about 2.11\,GHz is emitted at time $t_1$ by a DSN antenna and received onboard the spacecraft at time $t_2$. 
The frequency is multiplied by a constant ratio of 240/221 by a transponder and sent back to a ground antenna where it was received at time $t_3$. 
The Doppler shift, that is the difference between the up- and down- frequencies, is called a 2-way observable if the two antennas are the same, 
a 3-way observable if they are different. 
In fact, the ODF observable is the average of the Doppler shift over a time span called compression interval \citep{moyer}. 
The ODF format, described in \citep{wackley}, also contains the compression time, the date of the middle of the compression interval, 
the emitted frequency and the receiving and transmitting DSN antenna identifiers.

To analyse the ODF data, a software called ODYSSEY has been developed at the Observatoire de la C\^ote d'Azur within a collaboration with Onera. ODYSSEY stands for ``Orbit Determination and phYsical Studies in the Solar Environment Yonder''. One of the motivation for the development of this software is the simulation of the expected effects of the Solar System Odyssey project.
It is basically an interplanetary trajectory determination software. 
It performs numerical integration in rectangular coordinates of dynamical equations to propagate the position and velocity of the spacecraft 
and variational equations to propagate the sensitivity of the position and velocity with respect to the initial conditions of position 
and velocity and other parameters to be fitted. 
The numerical integration uses the Adams-Moulton-Cowell algorithm \citep{henrici} at an order 10. 
The use of this algorithm enables the direct integration of second order equations.

The values of the parameters to be estimated are obtained through a best fit procedure. 
The difference between measured observables $O$ and the calculated ones $C$, which depend on the parameters $E$ 
to be fitted, is linearized in terms of variations of $E$. 
For N values of the observable, this corresponds to a linear system of N equations with as many variables 
as there are terms in $E$. This system is then solved in ODYSSEY with classical iterative least squares analysis. 

Maneuvers are taken into account as increments of velocity $\delta V$ along the three directions. 
The dates of the maneuvers are provided by JPL and their amplitudes estimated as parameters in the best fit analysis.
Constraints are imposed on the maximum values of the estimated $\delta V$ in non radial directions.

The dynamical model to compute the motion of the spacecraft includes gravitational attraction by the main bodies 
of the solar system and direct radiation pressure. 
The motion of the spacecraft is computed using non relativistic gravitational equations,
which are known to be sufficient at the considered level of accuracy \citep{markwardt}. 
For the solar radiation pressure we use the same model as in \citep{anderson} with 
the reflectivity coefficient $k_r=1.77$, the mass $m=259$\,kg of the spacecraft and a cross sectional area 
$S=5.9$\,m$^2$ (orientation of the antenna supposed here to be constant).
The accuracy of this model is not critical because the acceleration due to the solar radiation pressure has decreased to less than 
0.2\,nms$^{-2}$ for Pioneer 10 at the considered distances.
Other known standard perturbations have been found to be negligible for Pioneer at the considered period
\citep{anderson}.

In its present status, the data analysis does not take into account the detailed thermal models of the spacecraft, 
currently under study by different groups. 
These models are expected to produce a slowly evoluting radiation force due to heat dissipation from the Radioisotope Thermoelectric Generators (RTG); this force should appear as a part of
the anomalous  secular acceleration to be found below. 
Improved analysis of this important issue will only be possible when accurate enough models of 
a time dependent radiative budget will be available. 

The dynamical equations are integrated in the Barycentric Celestial Reference System (BCRS); The reference time scale is the Barycentric Coordinate Time (TCB). 
Positions of terrestrial stations are expressed in the International Terrestrial Reference System (ITRS); the reference time scale is the Coordinated Universal Time (UTC).
The transition between ITRS and BCRS on the one hand, and TCB and UTC on the other hand, are performed according to 2003 International Earth Rotation Service (IERS) conventions. 
The positions of celestial bodies are obtained from DE 405 ephemeris from JPL in BCRS with a reference time scale which is refered to as ``$T_{eph}$'' and which is similar to the Barycentric Dynamical Time \citep{standish}.  

Special efforts have been devoted in the development of ODYSSEY for handling the calculation of the ODF observable. 
In a first step, the perturbations of the  round-trip light time are not taken into account. 
The ODF observable, that is the average of the Doppler shift over the compression time, is computed 
through a numerical approximation using the 4-points Simpson method. 
For the available Pioneer data, the compression time does not exceed 1980\,s.
The accuracy of the 4-points Simpson method in this case is evaluated to be 0.4\,mHz. 

The instantaneous Doppler shift is calculated in terms of velocities of the endpoints,
evaluated at the event times $t_1$, $t_2$ and $t_3$ \citep{markwardt}.
As only $t_3$ is provided in ODF, $t_2$ and $t_1$ have to be determined, which
is done iteratively using the light time equation
 \begin{eqnarray}
 \label{eq:1}
 t_{2}=t_{3}-\frac{r_{23}(t_2,t_3)}{c}-\delta \ell_{23}(t_2,t_3) \;,\quad 
 t_{1}=t_{2}-\frac{r_{12}(t_1,t_2)}{c}-\delta \ell_{12}(t_1,t_2)
 \end{eqnarray}
where $\delta \ell_{12}$ and $\delta \ell_{23}$ are respectively the Shapiro time delay affecting the up- and down-link signals. 

The spin frequency involved in Doppler shift estimation is provided by JPL. 
The ODF observable comes from JPL already corrected for data after the 17$^{\textrm{th}}$ of July 1990. 
Before this date, the spin correction is done by our software, taking the closest available value in the provided time series. 
As the maximal difference between two successive spin values in the file is 0.2\,mHz, the accuracy of this method is considered 
as sufficient.
  
In the second step of the computation of the observable, the perturbations which affect the propagation of the tracking signal 
are taken into account. The Shapiro time delay and the solar corona effect are modeled as in \citep{anderson}.
For the determination of the electron density necessary to compute ionospheric effect, two models have been implemented,
the International Reference Ionosphere (IRI) 2007 \citep{bilitza} and the Parametrized Ionospheric Model (PIM) \citep{daniell}.

Two models have been implemented in ODYSSEY for the mapping functions of the tropospheric effect, the Niell Mapping Functions (NMF) 
\citep{niell} and the Global Mapping Function (GMF) \citep{boehm2_gmf}.  

All these perturbations are modeled as delays affecting the signal (except for the Shapiro delay, their effect in equation (\ref{eq:1}) is however negligible). 
They are written as added propagation lengths, $\delta \ell_{12}$ and $\delta \ell_{23}$ for the up-
and down-links. Their effect on the ODF observable is therefore given by the difference
of these quantities between the ``start'' and ``stop'' of the compression interval \citep{moyer}.

\section{Confirmation of the existence of a constant anomaly}
Our first aim was to study the secular Pioneer anomaly reported by Anderson et al. \citep{anderson}. 
To this aim, we performed a best-fit with a constant anomalous acceleration $a_P$ exerted on the probe. 
The acceleration was centered on a point chosen at the Sun, the Solar System Barycentre (SSB) or the Earth. 
The results obtained with these 3 possibilities were not distinguishable from each other, due to
the large distance of the probe to the center of the solar system. 
From now on, we consider the center to be the Sun. 

The initial conditions as well as the three components of each maneuver are also fitted. 
The components of the maneuvers are fitted in the RTN (Radial Transverse Normal) frame 
and the transverse and normal amplitudes are constrained to be inferior to $0.2$\,ms$^{-1}$. 
The IRI 2007 model for ionospheric correction and GMF for tropospheric correction are used. 
Points with an elevation inferior to 20$^{\circ}$ are rejected so as to limit the effect
of imperfections of atmospherical models.
Outliers are also rejected when their difference with the expectation exceeds 100\,Hz at the first iteration 
and 6$\sigma$ at the following iterations with $\sigma$ the standard deviation of the residuals at this iteration.

\begin{figure}[h!]
 \begin{minipage}[t]{.48\linewidth}
  \centering\epsfig{figure=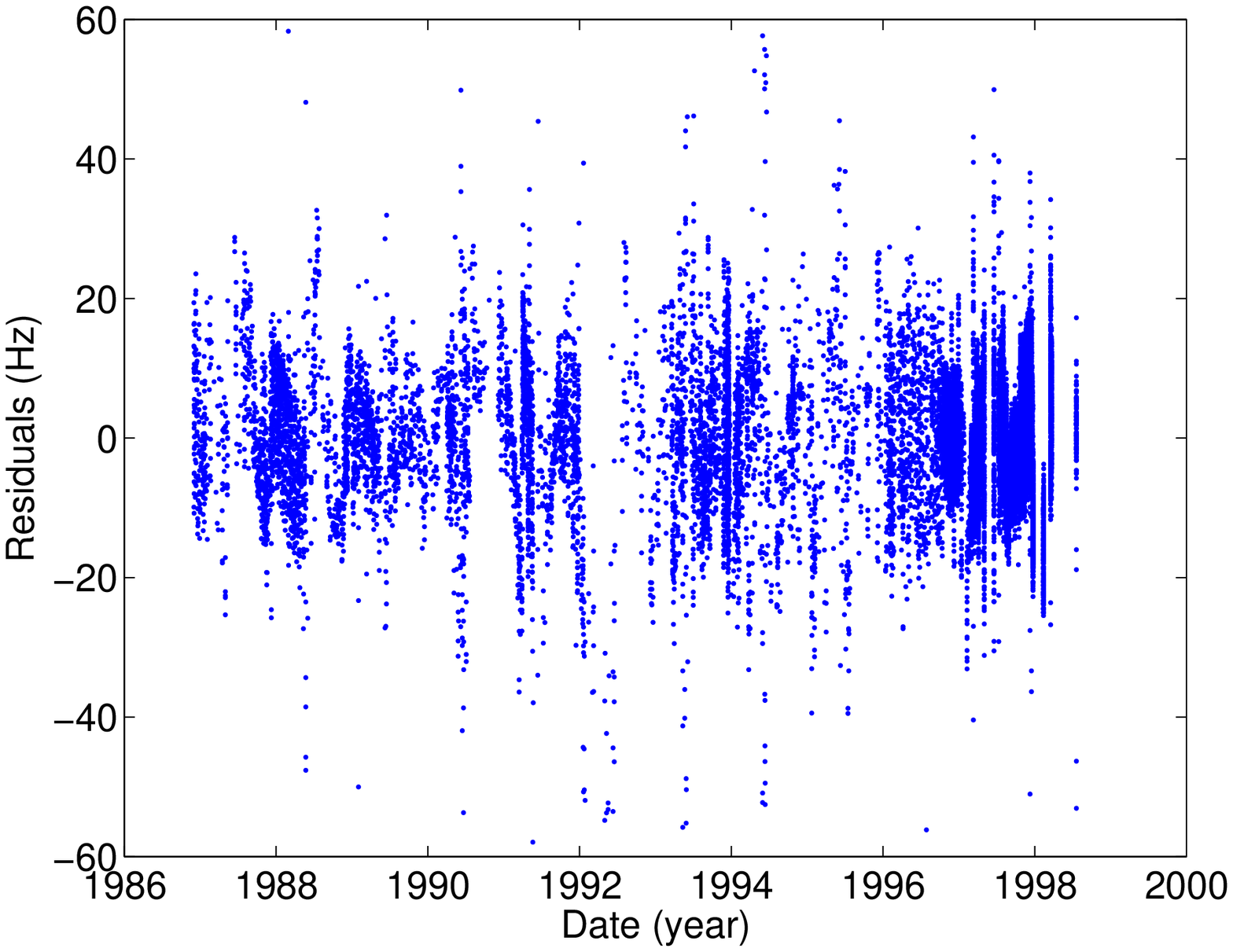,width=\linewidth}
  \caption{Best fit residuals of the Doppler tracking data of Pioneer 10 with an anomalous acceleration $a_P$.\label{figure1}}
 \end{minipage} \hfill
 \begin{minipage}[t]{.48\linewidth}
  \centering\epsfig{figure=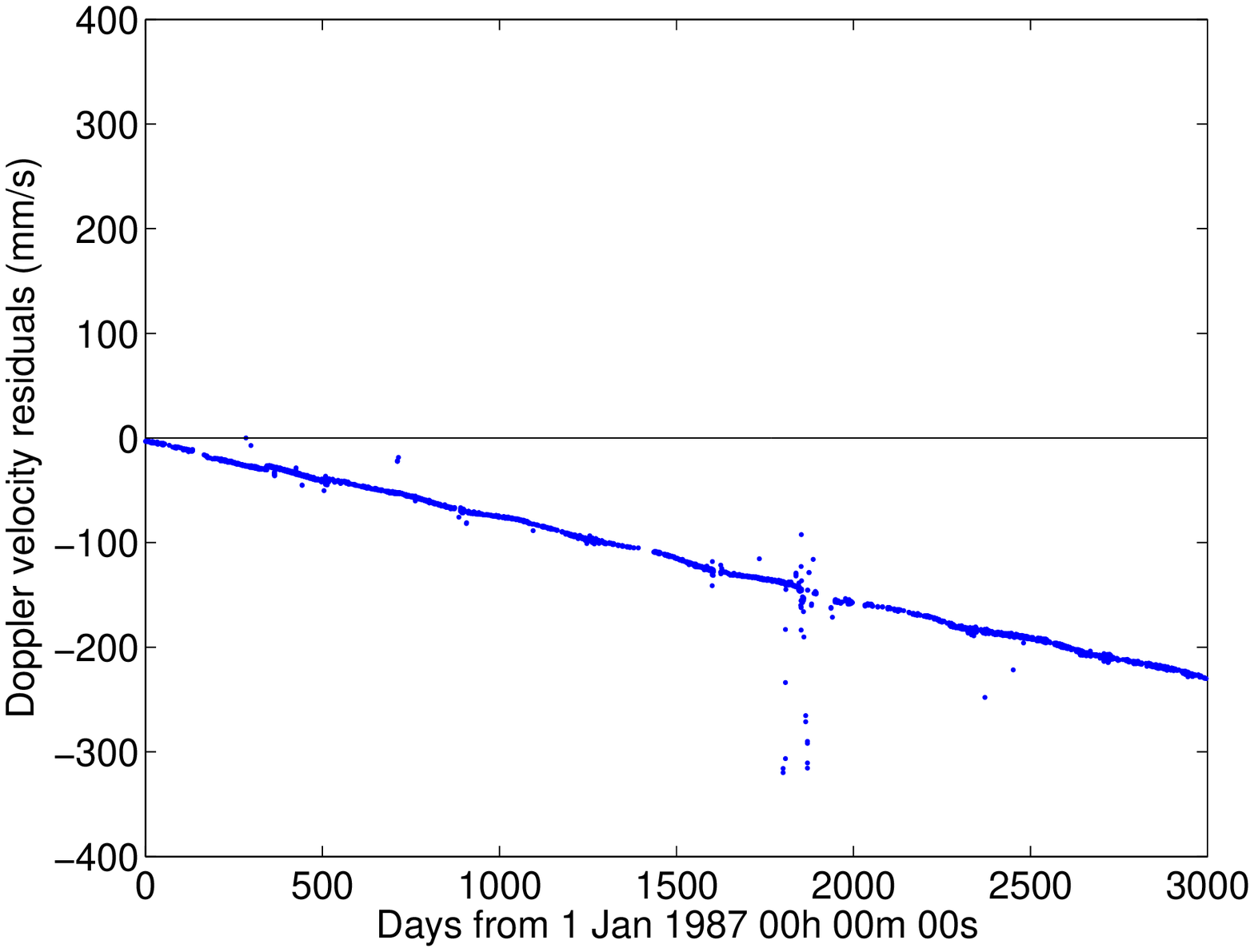,width=\linewidth}
  \caption{Reconstruction of an anomalous contribution, as in Fig. 8 of \citep{anderson}; 1\,mms$^{-1}$ corresponds to 15.4\,mHz.\label{figure2}}
 \end{minipage}
\end{figure}

The analysis performed with the software ODYSSEY confirms that a better fit is obtained with a constant sunward acceleration. 
The value estimated by ODYSSEY for the anomalous acceleration is $ a_P=0.84\pm0.01$\,nms$^{-2}$ with the formal error given at 1$\sigma$.
The postfit residuals show a standard deviation of 9.8\,mHz, which is largely improved with respect to a fit without $a_P$. 
These residuals are shown on Fig.~\ref{figure1}. 

On the Fig.~\ref{figure2} is given another representation agreeing with figure 8 of Anderson \citep{anderson}. 
This figure is reconstructed by the following procedure: the postfit residuals are expressed as $O-C(X_0^{opt},\dot{X}_0^{opt},M^{opt},a_P)$ 
where $X_0^{opt}$, $\dot{X}_0^{opt}$ and $M^{opt}$ are the optimized values for initial position, initial velocity and maneuver amplitudes; 
then the Fig.~\ref{figure2} represents the quantity $O-C(X_0^{opt},\dot{X}_0^{opt},M^{opt},0)$ where the anomalous acceleration has been
nullified; this representation highlights the need of the constant acceleration to reduce the residuals. 

\begin{figure}[h!]  
\centering 
\includegraphics[width=7cm]{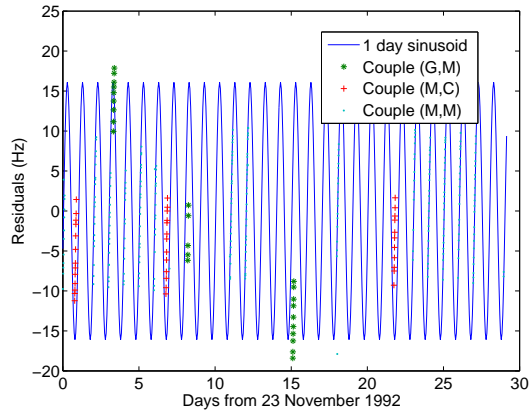}
\caption{Best fit residuals of the Doppler tracking data of Pioneer 10, for a one-month period near opposition. Different symbols or colors refer
to different couples of stations (see the text for details).} 
\label{figure3} 
\end{figure}

It can be emphasized that the level of the residuals on Fig.~\ref{figure1} is higher than the measurement noise. 
It is also clear on the figure that the postfit residuals do not correspond to a white gaussian noise. 
In order to highlight the existence of systematic structures in the residuals, we zoom 
on a time interval corresponding to the period from 23 November 1996 to 23 December 1996
where Pioneer 10 was on opposition (Sun, Earth and Pioneer 10 aligned in this order).
The data set is thus less affected by solar plasma and it shows daily variations of the residuals 
(see Fig.~\ref{figure3}). 
This figure clearly shows the existence of daily variations in the residuals, which are  emphasized by distinguishing
the data according to different couples of emission/reception antennas.
Different symbols or colors refer to different couples of stations, with M standing for Madrid (DSN antenna 63), 
G for Goldstone (DSN antenna 14), C for Canberra (DSN antenna 43).
The full line is a daily sinusoid fitted to the residuals of the dominant data (M,M).

\section{Study of the periodic variations of the Pioneer Anomaly} 

In order to characterize the periodic variations of the residuals, we have performed a spectral analysis of the postfit residuals. 
As the Pioneer data points are not evenly distributed, the commonly used Fourier transform is not appropriate for the analysis. 
We have used the software SparSpec which estimates frequency distributions in time series \citep{sparspec}. 
The result of this spectral analysis is shown on Fig.~\ref{figure5}.

The presence of significant periodic terms is clear at the periods measured with respect to a day = 86400\,s:
\begin{eqnarray}
f_1&=&0.9974\pm 0.0004 \,\textrm{day} \nonumber \\ 
f_2&=&\frac{1}{2}(0.9972\pm 0.0004) \,\textrm{day} \nonumber \\
f_3&=&189\pm 32 \,\textrm{day}. \nonumber
\end{eqnarray}
As 0.9972 day = 1.0 sidereal day, these periods are consistent with variations on one sidereal day, half a sidereal day, 
and half a year. 

The presence of diurnal and seasonal variations in the residuals has also been reported by Anderson et al. \citep{anderson}. 
It has to be noted that these specific periods are unlikely to be due to anything related to the spacecraft
or its environment. 
Errors in the atmospherical models would induce daily variations.
As these effects depend on the conditions at the stations, such errors would be expected at solar day rather than at sidereal day. 
Anderson et al. \citep{anderson} propose modeling errors such as in the case of Earth's ephemeris, 
the orientation of Earth's spin axis or the station's coordinates. However, these parameters are strongly constrained by other observational methods and it seems difficult to change them enough to explain the periodic anomaly.

\begin{figure}[h]  
\centering 
\includegraphics[width=7cm]{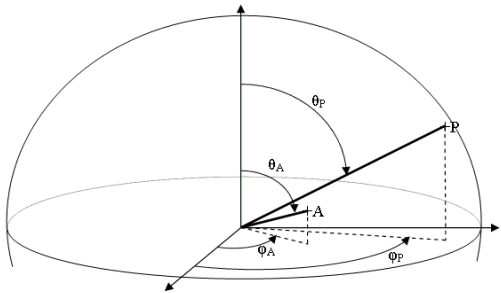}
\caption{Definition of the angle $\varphi$.} 
\label{figure4} 
\end{figure}

\begin{figure}
 \begin{minipage}[t]{.48\linewidth}
  \centering\epsfig{figure=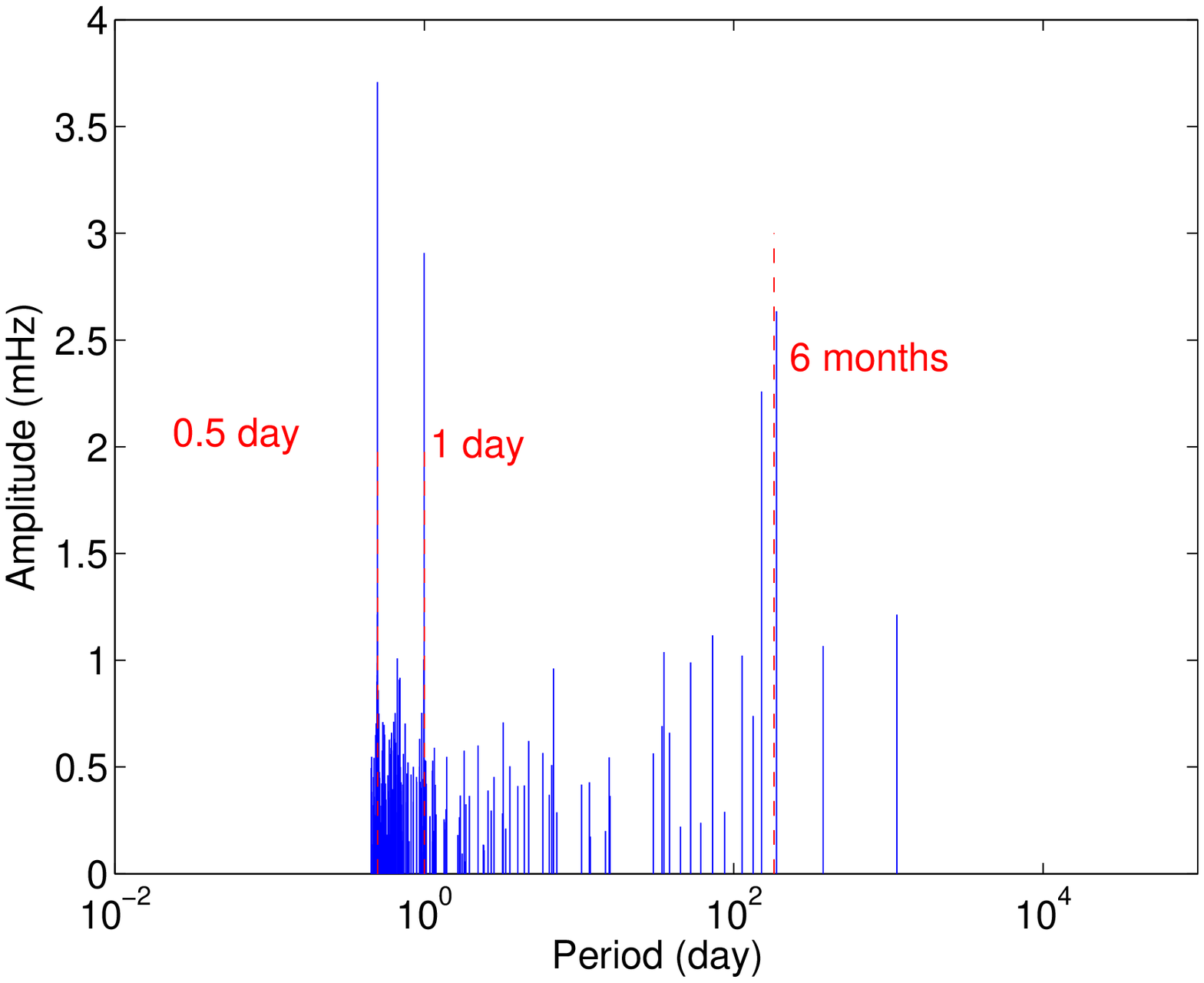,width=\linewidth}
  \caption{SparSpec analysis of the residuals from the fit with a constant acceleration.\label{figure5}}
 \end{minipage} \hfill
 \begin{minipage}[t]{.48\linewidth}
  \centering\epsfig{figure=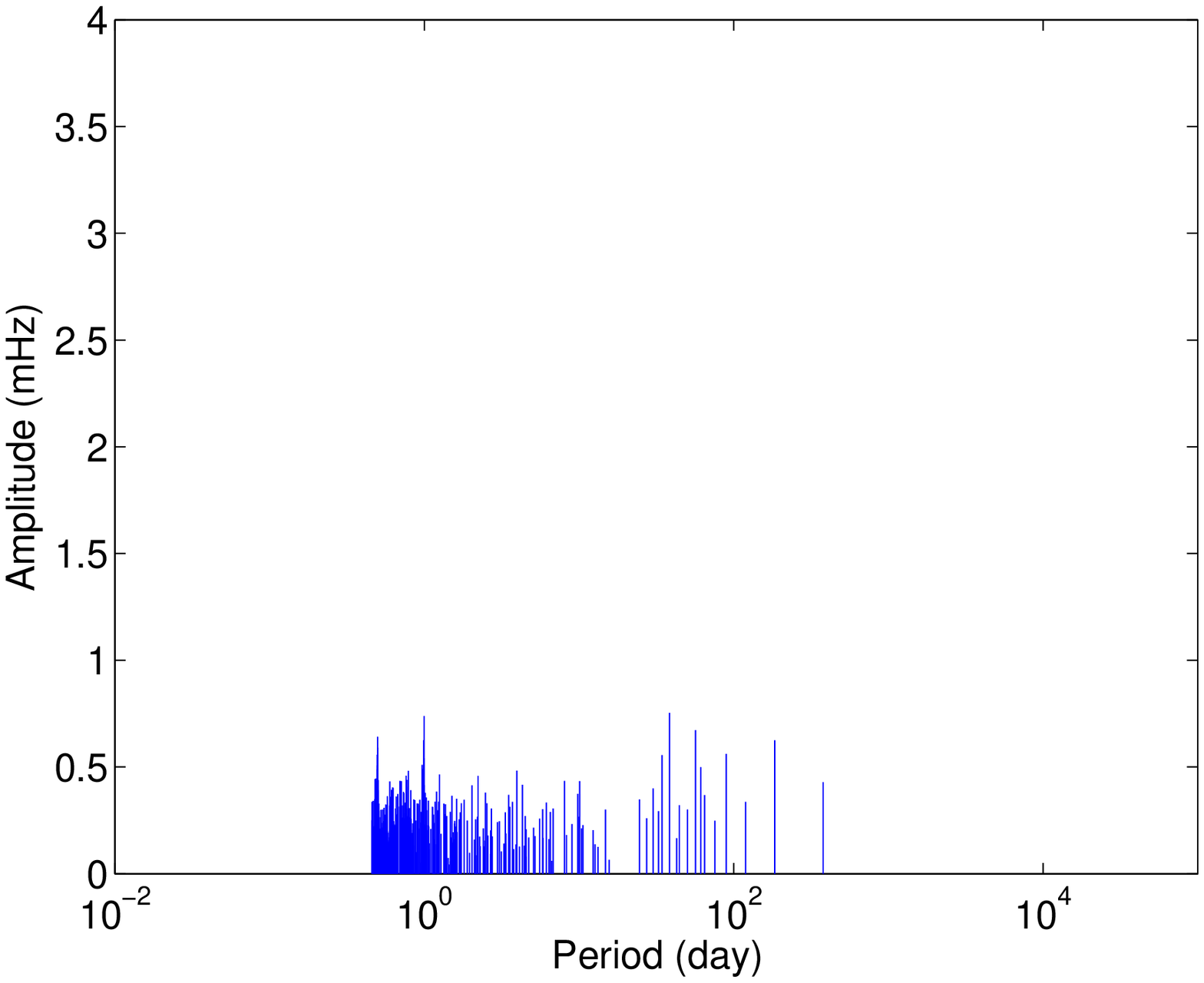,width=\linewidth}
  \caption{SparSpec analysis of the residuals from the fit with a constant acceleration and periodic terms.\label{figure6}}
 \end{minipage}
\end{figure}

The main motivation of the present paper is to test an alternative explanation where
some perturbation would modify the propagation of the tracking signal along the path
from the Earth antenna and the spacecraft. 
The idea is to represent such a perturbation, whatever its origin, as a function of the angle $\varphi$ defined as the difference between the Earth Antenna (A) azimuthal angle and the Pioneer (P) azimuthal angle : $\varphi = \varphi_P-\varphi_A$ (see Fig.~\ref{figure4}). 
The main interest of this geometrical model is that it should simultaneously account for the orbital movement of the 
Earth around the Sun and the diurnal rotation of the Earth. 

As this perturbation is supposed to be periodic, it will be represented by a few Fourier coefficients
\begin{equation}
 \label{eq:2}
\Delta f=\sum_n \left(\upsilon_n(\cos(n\varphi_u)+\cos(n\varphi_d))+\upsilon^\prime_n(\sin(n\varphi_u)+\sin(n\varphi_d)) \right)
\end{equation} 
Here $\varphi_u$ and $\varphi_d$ are the angles $\varphi$ evaluated on the up- and down-links.
Several tests have been performed with different numbers of coefficients $\upsilon_n$ and $\upsilon^\prime_n$ fitted in addition to 
the constant anomalous acceleration model. 
The best results have been obtained with superpositions of functions with $n=1$ and $2$.
The addition of higher order coefficients has not modified significantly the residuals.
To summarize, the new fit is identical to the previous one but for the addition
of the following modification of the Doppler signal:
\begin{equation}
\begin{split}
 \label{eq:3}
\Delta f=\upsilon_1(\cos(\varphi_u)+\cos(\varphi_d))+\upsilon^\prime_1(\sin(\varphi_u)+\sin(\varphi_d))\\
+\upsilon_2(\cos(2\varphi_u)+\cos(2\varphi_d))+\upsilon^\prime_2(\sin(2\varphi_u)+\sin(2\varphi_d))
\end{split}
\end{equation}  
 
This model results in a spectacular improvement of the best fit residuals, with the standard deviation 
reduced by a factor of the order of two.
Precisely, the standard deviation of the residuals which was 9.8\,mHz without the periodic terms
has been decreased to 5.5\,mHz. 
The values of the fitted anomalous parameters are reported in table~\ref{table1},
with two different ionospheric models.     

The robustness of the results with respect to other modifications of the
procedure is evaluated by the few tests which follow.
Using the IRI model and selecting data according to minimal elevation 
or outlier definition, we obtain the results shown in the tables~\ref{table2} and~\ref{table3}. 
These tables show that the best fit results are robust enough, 
except for the amplitude $\upsilon_1$, when details of the procedure are changed.
In particular, the anomalous secular acceleration is not very sensitive
to these changes and it remains roughly the same that in the previous
best analysis without modulated terms.
For the parameters other than $\upsilon_1$, the uncertainty in the determination 
has to be estimated from the variations of the results when the details are changed,
and certainly not from the formal errors given for a given fit which are smaller.

For the amplitude $\upsilon_1$, which is clearly less stable than the other parameters,
it is better to consider that it is not determined by the best fit procedure.
This feature can plausibly be understood from the following argument.
The direction from the Sun to Pioneer 10 varies only slowly, because of the large
distance of the probe. It follows that the periodic terms contain essentially variations
at one year and one sidereal day for $n=1$, at half a year and half a sidereal day for $n=2$. 
The fit of the initial conditions also induces terms at the same periods.
In particular, a change of the initial conditions may easily produce variations
masking modulated terms \citep{courty,jaekel}.
This is probably the reason why the yearly period, which corresponds to a large potential amplitude,
was in fact not detected in the spectral analysis of the best fit residuals with only a secular anomaly. 

Now the best fit with modulated terms tell us a different story. 
The modulated terms are looked for in a dedicated best fit procedure and 
they are unambiguously found to differ from zero.
The values of $a_P$, $\upsilon_1^\prime$, $\upsilon_2$ and $\upsilon_2^\prime$
seem to have robust estimates, while $\upsilon_1$ is less stable. The analysis of the correlation shows a large value between $\upsilon_1$ and the initial conditions (0.814 compared to 1 for a total correlation). There is an interdependance between the estimated values of $\upsilon_1$ and the initial conditions. This large correlation reflects the phenomenon described in the previous paragraph explaining the instability of $\upsilon_1$ estimation.

An even more impressive demonstration of the improvement of the data analysis drawn
by the inclusion of modulated terms comes from the spectral analysis of the residuals.
This spectral analysis is represented on Fig.~\ref{figure6} for the best fit with
modulated terms. It is drawn intentionnally at the same scale as Fig.~\ref{figure5}
so that one can easily notice the global reduction of the main peaks in the spectrum
as well as all the secondary peaks. 

\section{Conclusion}
In the present paper, we have reported the first results of our re-analysis 
of the Pioneer 10 Doppler data for the 1986 to 1998 time span.
The data are the same that were already analysed in \citep{anderson,markwardt,olsen}
but they have been dealt with by using a dedicated software ODYSSEY developed to this purpose. 
The improvement of the data fit with a constant anomalous acceleration exerted on Pioneer 10 
has been confirmed by this new data analysis. Its magnitude is compatible with
that reported by Anderson et al. 

The paper has then been focused on the study of periodic terms in the residuals, 
which were already noticed in previous studies and are clearly revealed using spectral
analysis. 
The main new result of the paper is that a large part of these diurnal and seasonal
anomalies may be captured in a simple geometrical model where the light time 
on the tracking path is modified in a manner depending only on the azimuthal angle $\varphi$
between the Sun-Earth and Sun-probe lines. 
This geometrical model could represent in a simple way the physical effects expected on
light propagation in some metric extensions of general relativity which have been
studied as potential candidates for the explanation of the secular Pioneer anomaly \citep{jaekel,jaekel2}. The search for modulated anomalies might even be used to distinguish
those candidates from other which affect much less the propagation of tracking signals
\citep{moffat}. However, the results of the paper cannot be considered as pointing to a particular possible explanation of the anomaly. 
At the present stage of the data analysis, similar effects could for example be obtained 
through a mismodeling of the solar corona model. 

Anyway, considering the modulated anomalies has allowed us to reduce by a factor of the order of two 
the standard deviation of the residuals.
The results of the best fit have been found to be robust enough with respect to changes in 
the tropospheric or ionospheric model, as well as variations of the details of the best fit procedure.
The most impressive output of the new analysis is given by the SparSpec analysis of the best fit residuals 
which shows a reduction of all periodic structures which were present in the residuals of the best fit 
without modulated terms. It is worth emphasizing that this has been done by considering only a
function of the geometrical angle $\varphi$, and not by using different explanations for the 
diurnal and seasonal anomalies.
This suggests that the new analysis constitutes a richer characterization of the Pioneer data,
now involving not only a secular acceleration but also modulated terms, which will have to be 
compared with any, existing as well as future, possible explanation of the anomaly.

\section*{Acknowledgements}
This work has benefited of discussions with a number of people involved in the international collaborations
devoted to the investigation of the Pioneer data. 
Special thanks are due to S.G. Turyshev (NASA JPL) for having led the ISSI team \citep{issi} during
which the ODF used in the seminal paper \citep{anderson} were made available to the members of the collaboration. In particular, the development of ODYSSEY software has benefited from fruitful discussions with Pierre Exertier (OCA). We are also grateful for discussions with the members of this collaboration and especially with P. Touboul and B. Foulon (ONERA) and the members of the french collaboration GAP ({\em Groupe Anomalie Pioneer}) \citep{gap}. 
Another special thanks is due to CNES for its support of the GAP.

\clearpage
\begin{table}[h!]
\caption{Results of the best fit with periodic terms in the signal, with two different ionospheric models. } 
\centering 
\begin{tabular}{c c c c} 
\hline\hline 
TEC Model&IRI 2007&PIM \\ 
\hline 
$a_P$ (pms$^{-2}$)&-836$\pm$1&-836$\pm$1\\ 
\hline  
$\upsilon_1$ (mHz)&124.3$\pm$9.3&141.6$\pm$9.3\\ 
\hline  
$\upsilon^\prime_1 $ (mHz)&-125.3$\pm$0.6&-127.3$\pm$0.6\\ 
\hline  
$\upsilon_2$ (mHz)&2.7$\pm$0.2&3.0$\pm$0.2\\ 
\hline 
$\upsilon^\prime_2$ (mHz)&-4.8$\pm$0.1&-4.9$\pm$0.1\\ 
\hline  
Residuals (mHz)&5.5&5.5\\
\hline 
\end{tabular} 
 
\label{table1}
\end{table}

\begin{table}[h!]
\caption{Results of the best fit with periodic terms in the signal, 
when varying the minimal evaluation for the considered data. The initial number of points is 19805.}  
\centering 
\begin{tabular}{c c c c c}  
\hline\hline 
Min. elevation&0$^\circ$&10$^\circ$&20$^\circ$&30$^\circ$ \\  
\hline  
$a_P$ (pms$^{-2}$)&-844$\pm$1&-843$\pm$1&-836$\pm$1&-826$\pm$1\\ 
\hline 
$\upsilon_1$ (mHz)&86.4$\pm$9.3&98.0$\pm$9.2&124$\pm$9.3&95$\pm$10.5\\ 
\hline  
$\upsilon^\prime_1 $ (mHz)&-111.8$\pm$0.6&-119.5$\pm$0.7&-125$\pm$0.6&-127$\pm$0.6\\ 
\hline 
$\upsilon_2$ (mHz)&2.1$\pm$0.2&2.3$\pm$0.2&2.7$\pm$0.2&2.1$\pm$0.2\\ 
\hline 
$\upsilon^\prime_2$ (mHz)&-4.7$\pm$0.1&-4.7$\pm$0.1&-4.8$\pm$0.1&-4.6$\pm$0.1\\ 
\hline 
Nr of eliminated points&789&911 &2996&8584\\ 
\hline 
Residuals (mHz)&6.0&5.9 &5.5&5.1\\ 
\hline
\end{tabular} 
\label{table2}
\end{table}

\begin{table}[h!]
 \caption{Results of the best fit with periodic terms in the signal, 
when varying the outlier criterion at N$\sigma$. The initial number of points is 19805.}  
\centering  

\begin{tabular}{c c c c }  

\hline\hline 
Outliers at $N\sigma$&N=3&N=6&N=8 \\ 
\hline 
$a_P$ (pms$^{-2}$)&-806$\pm$1&-836$\pm$1&-848$\pm$1\\ 
\hline 
$\upsilon_1$ (mHz)&197$\pm$7.2&124$\pm$9.3&102$\pm$10.0\\ 
\hline  
$\upsilon^\prime_1 $ (mHz)&-126$\pm$0.5&-125$\pm$0.6&-124$\pm$0.7\\ 
\hline
$\upsilon_2$ (mHz)&3.4$\pm$0.1&2.7$\pm$0.2&2.6$\pm$0.2\\ 
\hline  
$\upsilon^\prime_2$ (mHz)&-5.0$\pm$0.09&-4.7$\pm$0.1&-4.7$\pm$0.1\\ 
\hline  
Nr of eliminated points&3879&2996 & 2933\\ 
\hline  
Residuals (mHz)&4.16&5.5 &5.9 \\ 
\hline  
\end{tabular} 
\label{table3}
\end{table}

\end{document}